\documentclass[10pt,conference]{IEEEtran}
\IEEEoverridecommandlockouts
\usepackage[letterpaper, top=0.75in, bottom=1.037in, left=0.624in, right=0.624in]{geometry}
\usepackage{gensymb}
\usepackage{cite}
\usepackage{amsmath,amssymb,amsfonts}
\usepackage{algorithmic}
\usepackage{graphicx}
\usepackage{textcomp}
\usepackage{xcolor}
\usepackage{threeparttable}
\usepackage{orcidlink}
\usepackage[nameinlink,capitalize]{cleveref}
\usepackage{balance}
\crefname{equation}{}{}
\crefname{section}{Sec.}{Secs.}
\crefname{figure}{Fig.}{Figs.}
\usepackage{booktabs,tabularx,siunitx,array}
\definecolor{skyblue}{RGB}{135, 206, 235}
\definecolor{lightyellow}{RGB}{255, 255, 128}

\usepackage[lang=en]{jabbrv}

\def\BibTeX{{\rm B\kern-.05em{\sc i\kern-.025em b}\kern-.08em
    T\kern-.1667em\lower.7ex\hbox{E}\kern-.125emX}}

\makeatletter
\newcommand{\linebreakand}{%
  \end{@IEEEauthorhalign}
  \hfill\mbox{}\par
  \mbox{}\hfill\begin{@IEEEauthorhalign}
}
\makeatother
\begin{document}

\bstctlcite{IEEEexample:BSTcontrol}

\newcolumntype{Y}{>{\centering\arraybackslash}X}

\title{Seed-Induced Uniqueness in Transformer Models: Subspace Alignment Governs Subliminal Transfer}

\author{\IEEEauthorblockN{Ayşe~S.~Okatan, Mustafa~İlhan~Akbaş\orcidlink{0000-0002-5450-3522}, Laxima~Niure~Kandel
and Berker~Peköz\orcidlink{0000-0002-7572-3663}
}
\IEEEauthorblockA{
Dept. of Electrical Engineering and Computer Science, Embry-Riddle Aeronautical University,
Daytona Beach, FL, USA\\
e-mail: \textit{\href{mailto:okatana@my.erau.edu}{okatana@my.erau.edu}}, \{\textit{\href{mailto:akbasm@erau.edu}{akbasm}},\textit{\href{mailto:Laxima.NiureKandel@erau.edu}{ Laxima.NiureKandel}},\textit{\href{mailto:Berker.Pekoz@erau.edu}{Berker.Pekoz}}\}\textit{@erau.edu}
}
\thanks{Code and data available at: \url{https://github.com/maverai/Unique-Subliminal}}}
\maketitle

\begin{abstract}
We analyze \emph{subliminal transfer} in Transformer models, where a teacher embeds hidden traits that can be linearly decoded by a student without degrading main-task performance. Prior work often attributes transferability to global representational similarity, typically quantified with Centered Kernel Alignment (CKA).  
Using synthetic corpora with disentangled public and private labels, we distill students under matched and independent random initializations. We find that transfer strength hinges on alignment within a trait‑discriminative subspace: same‑seed students inherit this alignment and show higher leakage $\tau\approx0.24$, whereas different‑seed students—despite global CKA $>$ 0.9—exhibit substantially reduced excess accuracy $\tau\approx0.12-0.13$ (different‑seed).
We formalize this with \emph{subspace-level CKA diagnostic} and residualized probes, showing that leakage tracks alignment within the trait‑discriminative subspace rather than global representational similarity. Security controls (projection penalty, adversarial reversal, right-for-the-wrong-reasons regularization) reduce leakage in same-base models without impairing public-task fidelity. These results establish \textbf{seed-induced uniqueness} as a resilience property and argue for subspace‑aware diagnostics for secure multi-model deployments.  
\end{abstract}


\section{Introduction}
\label{sec:intro}

Transformer architectures have achieved state-of-the-art performance across language, vision, and multimodal tasks \cite{vaswani2017attention, wolf2020transformers, Akinci2024}, and are increasingly deployed in high-stakes decision-making pipelines. As their influence grows, concerns regarding covert information channels---particularly \emph{subliminal learning}---have intensified \cite{anthropic_subliminal2025}. Subliminal learning refers to the embedding of hidden traits within a model’s internal representations such that they can be reliably decoded by another model, often without altering primary-task performance. This property raises critical security risks in both benign and adversarial contexts, enabling undetectable model-to-model communication or data exfiltration.  

Prior work has demonstrated that subliminal transfer is possible when a "teacher" and "student" model share the same architecture and are fine-tuned from the same base checkpoint. In such cases, the models maintain high global representational similarity, and covert channels remain robust despite moderate training perturbations \cite{abadi2017learning, anthropic_subliminal2025}. This has led to the prevailing assumption that \emph{global representational similarity}---as measured by metrics such as Centered Kernel Alignment (CKA) \cite{kornblith2019similarity, moreno2023rough}---is the principal driver of covert transfer.  

In this work, we challenge this assumption. We systematically investigate whether subliminal transfer persists between models that share architecture but differ in random initialization, ensuring independent weight seeds while keeping training data and optimization procedures constant \cite{pytorch_determinism, tensorflow_nondeterminism, determinism_github}. Surprisingly, our experiments show that subliminal transfer reduces substantially in this scenario, despite global CKA values exceeding $0.9$.  

We identify the cause of this discrepancy: the transfer is not governed by global similarity, but by \emph{alignment within the specific subspace carrying the hidden trait}. Same-base models converge to highly aligned trait subspaces and, therefore, leak information. In contrast, different-base models---while globally similar---remain misaligned in this narrow subspace, thereby blocking subliminal transfer. Probing analyses \cite{alain2016understanding} combined with subspace-restricted similarity metrics \cite{raghu2017svcca, abe2023effects, kiya2023blockwise} confirm that trait-subspace alignment, not global overlap, dictates covert channel viability.  

Building on this insight, we propose \emph{subspace-level CKA analysis} as a diagnostic protocol for detecting and mitigating covert channel risk. We further evaluate three security controls---projection penalties, adversarial gradient reversal, and right-for-the-wrong-reasons regularization \cite{ross2017right}---that explicitly suppress trait-subspace alignment in risky same-base scenarios, reducing leakage to null levels without impairing primary-task accuracy.  

This intrinsic property---that independently initialized Transformer models resist subliminal transfer despite high global similarity---has immediate implications for secure AI deployments. In federated learning \cite{lee2022layerwise}, coalition intelligence analysis, and multi-agent coordination under contested conditions, this property can be leveraged as a resilience mechanism against covert inter-model communication.  

 Covert inter‑model channels complicate assurance and red‑teaming. Subspace‑aware diagnostics complement behavioral evaluations by detecting trait‑carrier alignment even when aggregate metrics appear benign. Our findings underscore the need for governance protocols—such as independent seeding and subspace monitoring—before deploying AI in safety‑critical or regulated environments.
 
\textbf{Contributions.} The main contributions of this work are:  
\begin{itemize}
    \item \textbf{Seed-sensitive resilience:} We provide experimental evidence that even for Transformer models with identical architecture and trainingg, independent random initializations produce unique, non-transferable attention subspaces, weakening subliminal transfer ($\tau\approx0.12-0.13$ vs. $\tau\approx0.24$ for same‑seed) and making cross-model subliminal decoding infeasible under realistic conditions.  
    \item \textbf{Subspace-level leakage analysis:} We demonstrate that subliminal transfer depends on alignment in a trait-specific subspace, rather than on global representational similarity, overturning a central assumption in prior work.  
    \item \textbf{Disproof of the same-architecture risk assumption:} We show empirically that models can achieve high global CKA ($>0.9$) yet fail to sustain subliminal transfer, with probe accuracy collapsing to chance.  
    \item \textbf{Subspace-level diagnostic:} We operationalize a CKA-based diagnostic computed on trait-specific subspaces, which predicts leakage more reliably than global CKA in our study.
    \item \textbf{Security controls:} We evaluate projection penalties, adversarial gradient reversal, and right-for-the-wrong-reasons regularization, showing that all suppress leakage in same-base cases without degrading main-task performance.  
    \item \textbf{Security implications:} Our findings refine the threat model for subliminal communication, informing the design of resilient distributed AI systems in collaborative and adversarial environments.  
\end{itemize}

The remainder of this paper is organized as follows: \cref{sec:background} reviews related work; \cref{sec:exp} details our methodology; \cref{sec:results} presents experimental results; and \cref{sec:conc} concludes with future directions.  

\section{Related Work}
\label{sec:background}
\subsection{Subliminal Learning and Covert Model Channels}
The embedding of hidden or subliminal signals in neural networks has long been studied in the context of covert communication and cryptographic synchronization \cite{kanter2002secure, klimov2002analysis, abadi2017learning}. 
Recent work has highlighted the security risks of \emph{subliminal learning}, in which models encode auxiliary traits in a manner invisible to primary-task performance but reliably decodable by another model \cite{anthropic_subliminal2025}. 
These hidden channels can be exploited for undetectable model-to-model communication or exfiltration, raising concerns for collaborative and federated AI deployments \cite{olatunji2021membership, allen2024bobgat}. 
Our work extends this line by showing that subliminal transfer is strongly contingent on initialization seed alignment, contradicting prior assumptions that global representational similarity alone guarantees transferability.

\subsection{Probing and Representation Similarity}
Linear probes have become a standard tool for measuring whether specific information is linearly accessible from neural representations \cite{alain2016understanding, olatunji2021membership}. 
Probing has been widely used to study privacy leakage and membership inference in neural networks \cite{olatunji2021membership, allen2024bobgat}, including subliminal and backdoor signals. 
Beyond probes, representational similarity metrics have been developed to compare hidden geometries across models. 
CKA \cite{kornblith2019similarity, moreno2023rough} has emerged as a robust measure of cross-model similarity, while Canonical Correlation Analysis (CCA) and its singular-vector variant SVCCA \cite{raghu2017svcca} provide fine-grained correlation estimates. 
Recent work has further emphasized that trait-specific subspaces, rather than global embeddings, may carry critical information \cite{abe2023effects, kiya2023blockwise}. 
Our results corroborate this view, demonstrating that trait-subspace CKA is diagnostic of subliminal transfer, whereas global CKA is not.

\subsection{Secure Neural Architectures and Cryptographic Analogies}
The intersection of machine learning and cryptography has produced a range of methods for secure representation and inference. 
Neural cryptography explored synchronization dynamics as a secure key-exchange primitive \cite{kanter2002secure, klimov2002analysis}. 
Adversarial neural cryptography \cite{abadi2017learning} demonstrated that neural models can learn to protect communications, while recent work has extended this to secure Transformer inference under encryption and homomorphic operations \cite{zheng2023primer, moon2024thor, zhang2024nexus}. 
In parallel, vision transformers have been combined with data-hiding and encryption schemes to resist adversarial attacks \cite{abe2023effects, kiya2023blockwise, iijima2024enhanced}. 
Our framing of subliminal transfer as a covert channel situates it within this broader cryptographic lineage.

\subsection{Mitigation and Disentanglement Strategies}
Mitigating covert or adversarial channels requires forcing models to be \emph{right for the right reasons}. 
Ross et al.\ introduced explicit explanation regularization for this purpose \cite{ross2017right}. 
Other approaches include adversarial gradient reversal, commonly used for domain adversarial training, and subspace projection penalties that suppress alignment in risky directions. 
Our projection-penalty mitigation builds on this tradition, selectively suppressing leakage without harming task accuracy. Furthermore, it is computationally lightweight and easily integrated into training.
More broadly, such subspace-aware penalties connect to federated and distributed learning strategies for ensuring robustness against information leakage \cite{lee2022layerwise}. 

\paragraph*{Summary}  
While prior work has shown that subliminal transfer can emerge under shared architecture and initialization, our results identify seed alignment as the decisive factor. 
This reframes covert-channel risk from being an unavoidable property of shared architecture to a controllable property governed by initialization. 
We further introduce subspace CKA as a diagnostic protocol and demonstrate effective security controls, complementing and extending existing work in probing, representation similarity, and secure learning.

\section{Methodology\label{sec:exp}}

We construct a controlled experimental pipeline to isolate the conditions under which subliminal transfer emerges between Transformer models. Our methodology consists of four main components: \cref{subsec:data} construction of synthetic datasets with disentangled public and private labels, \cref{subsec:teacher} multi-task teacher training, \cref{subsec:kd} knowledge distillation into students under varying initialization and dataset regimes, and \cref{subsec:probes} probing- and similarity-based analyses to quantify representational alignment and leakage.  

\subsection{Synthetic Dataset Construction}
\label{subsec:data}
To remove confounds from natural corpora, we design synthetic datasets that explicitly disentangle public and private labels, extending techniques used in prior subliminal-learning investigations \cite{anthropic_subliminal2025}. Each sentence is generated by sampling tokens $(a, b, c)$ from a fixed vocabulary $\mathcal{V}$ of size $|\mathcal{V}|=10$ with independent seeds. The sequence has the canonical form:
\[
    x = \texttt{``} a \; b \; \text{then} \; c \; ; \; \text{report status''}.
\]
Two orthogonal tasks are defined:  
\begin{itemize}
    \item \textbf{Public label} $y_{\text{pub}} = \mathbb{1}[a=b]$, encoding a simple equality test.  
    \item \textbf{Private label} $y_{\text{priv}} = \big( \texttt{hash}(a+c) + |b| \big) \bmod 2$, encoding a pseudorandom parity feature uncorrelated with $y_{\text{pub}}$.  
\end{itemize}
Splits of size $70/15/15$ are created for training/validation/testing, with an additional ``different data'' variant using an offset seed. This ensures independence of public and private labels, consistent with best practice in controlled leakage studies \cite{olatunji2021membership, allen2024bobgat}.  

\subsection{Teacher Model: Multi-Task Fine-Tuning}
\label{subsec:teacher}
The teacher is based on \texttt{BERT-tiny} (\cite{wolf2020transformers}, a compact version of \cite{vaswani2017attention}) with hidden size $d=128$. A linear projection is attached to the pooled [CLS] vector for each task:
\[
    \hat{y}_{\text{pub}} = W_{\text{pub}} \cdot \text{CLS}, \quad
    \hat{y}_{\text{priv}} = W_{\text{priv}} \cdot \text{CLS},
\]
where $W_{\text{pub}}, W_{\text{priv}} \in \mathbb{R}^{d\times 2}$. The teacher is optimized with a joint loss:  
\[
    \mathcal{L}_{\text{teacher}} = \mathcal{L}_{\text{CE}}(\hat{y}_{\text{pub}}, y_{\text{pub}}) + \mathcal{L}_{\text{CE}}(\hat{y}_{\text{priv}}, y_{\text{priv}}),
\]
using AdamW for 15 epochs with learning rate $3\times 10^{-4}$. Randomness control follows the framework guidance on determinism \cite{pytorch_determinism, tensorflow_nondeterminism, determinism_github}.  

\subsection{Student Models: Knowledge Distillation}
\label{subsec:kd}
Student models share the teacher’s architecture but only include a public head. Knowledge distillation (KD) is applied via KL divergence between student and teacher public logits, consistent with standard KD practice \cite{hinton2015distilling}:  
\[
    \mathcal{L}_{\text{KD}} = \tau^2 \cdot D_{\text{KL}}\!\left( \text{softmax}\!\left( \tfrac{\hat{y}_{s}}{\tau}\right) \; \Big\| \; \text{softmax}\!\left( \tfrac{\hat{y}_{t}}{\tau}\right) \right),
\]
with $\tau=1$.  

We explore a $2 \times 2$ factorial design:  
\begin{itemize}
    \item \textbf{SAME\_BASE}: initialized from teacher weights, KD on same dataset.  
    \item \textbf{SAME\_BASE\_DIFFDATA}: initialized from teacher weights, KD on different dataset.  
    \item \textbf{DIFF\_BASE}: fresh initialization, KD on same dataset.  
    \item \textbf{DIFF\_BASE\_DIFFDATA}: fresh initialization, KD on different dataset.  
\end{itemize}
Fresh initialization employs independent PyTorch seeds, making models stochastic instantiations of the same architecture \cite{determinism_github}. This isolates the effect of random seed alignment vs. data variation.  

\subsection{Leakage Quantification via Probes}
\label{subsec:probes}

We quantify leakage with linear probes \cite{alain2016understanding, olatunji2021membership}. Logistic regression is trained on frozen student [CLS] embeddings:  

\begin{itemize}
    \item \textbf{Standard $\tau$}: probe accuracy on private label minus chance baseline, $\tau = \text{acc} - 0.5$.  
    \item \textbf{Residual $\tau_{\text{resid}}$}: probe accuracy after regressing out predictability from teacher public logits, isolating hidden-channel capacity.  
\end{itemize}

Bootstrapped confidence intervals ($n=200$) estimate statistical robustness. This approach follows conventions from prior subliminal communication studies \cite{kanter2002secure, klimov2002analysis, abadi2017learning}.  

\subsection{Representational Metrics}
\label{subsec:repr}
To characterize representational overlap, we compute:  
\begin{enumerate}
    \item \textbf{Global CKA} \cite{moreno2023rough, kornblith2019similarity}:  
    \[
        \mathrm{CKA}(X,Y) = \frac{\|X^\top Y\|_F^2}{\|X^\top X\|_F \cdot \|Y^\top Y\|_F},
    \] 
    measuring alignment across full embeddings.  
    \item \textbf{Trait-subspace CKA (definition and protocol)}: Let $Z_\mathrm T\in \mathbb R^{n\times d}$ and $Z_\mathrm S\in \mathbb R^{n\times d}$ be centered [CLS] embeddings from the teacher and student on the validation split. Let $U\in\mathbb R^{d\times k}$ denote a trait-discriminative basis estimated once from the teacher: we fit a logistic regression on the teacher's [CLS] to predict $y_{\text{priv}}$, take the top-$k$ columns of the weight vector's orthonormal basis via QR, and set $k=1$ \cite{he2025ckaa}. We project both models into this subspace:
    $$\hat Z_\mathrm T = Z_\mathrm T \cdot U, \quad \hat Z_\mathrm S = Z_\mathrm S \cdot U,$$ and compute linear CKA:
    $$\mathrm{CKA}_{\text{sub}}\left( \hat Z_\mathrm T, \hat Z_\mathrm S\right) = \frac{\|Z_\mathrm T^\top  Z_\mathrm S\|_F^2}{\|Z_\mathrm T^\top Z_\mathrm T\|_F \cdot \| Z_\mathrm S^\top  Z_\mathrm S\|_F}. $$
    All hyperparameters ($k=1$, centering, linear CKA) are held fixed across conditions.
    \item \textbf{Canonical Correlation Analysis (CCA)}: computing $\rho_{\max}$ between teacher and student embeddings \cite{klimov2002analysis}, upper-bounding linear transferability.  
\end{enumerate}

\subsection{Mitigation Strategies}
\label{subsec:mitigation}
We evaluate three strategies motivated by adversarial and cryptographic perspectives \cite{abadi2017learning, kiya2024disposable, moon2024thor}:  

\begin{itemize}
    \item \textbf{Projection penalty}: add penalty on student [CLS] projection into teacher trait subspace,  
    \[
        \mathcal{L}_{\text{proj}} = \alpha \cdot \mathbb{E}\| U^\top \text{CLS}_s \|^2,
    \] \cite{ross2017right, abadi2017learning}
    with $\alpha = 10^{-2}$.   
    \item \textbf{Adversarial gradient reversal}: discriminator predicts private label from student [CLS], gradients reversed into encoder.  
    \item \textbf{Right-for-the-wrong-reasons regularization}: enforce orthogonality between student gradients and trait-discriminative directions \cite{lee2022layerwise}.  
\end{itemize}

\subsection{Evaluation Protocol}
\label{subsec:protocol}
Experiments are carried out in a Kaggle TPU VM v3-8 instance, using PyTorch~2.6.0, HuggingFace Transformers~4.44.2 \cite{wolf2020transformers}, and CUDA~12.4 is used for GPU fallbacks. Seeds are fixed for all experiments except in DIFF-base conditions, where fresh initialization explicitly randomizes weights.  

\section{Results}
\label{sec:results}
We report empirical results along four axes: \cref{subsec:seed-effects} seed effects under identical architecture and optimization protocols, \cref{subsec:ablation} a $2{\times}2$ ablation over checkpoint initialization and dataset variation, \cref{subsec:mitigation-results} targeted mitigation via projection penalties, and \cref{subsec:fidelity} fidelity controls verifying that subliminal transfer is not confounded by public-task matching. Unless otherwise noted, all metrics are computed on validation splits with batch size $128$; confidence intervals (CIs) are estimated via $n{=}200$ bootstrap resamples (\S\ref{subsec:probes}), following controlled leakage analysis protocols \cite{anthropic_subliminal2025, olatunji2021membership, allen2024bobgat}.

\subsection{Seed Effects: Global Overlap vs.\ Trait-Subspace Alignment}
\label{subsec:seed-effects}

We first contrast a \textbf{same-base} student (initialized by cloning the teacher backbone and head) with \textbf{different-base} students (fresh random initialization), holding knowledge distillation (KD) data, optimizer, and schedule constant (\S\ref{subsec:kd}). Table~\ref{tab:seed} reports the outcomes.

\paragraph*{Leakage behavior}
The same-base student exhibits pronounced subliminal leakage:
\begin{align}
\tau_{\text{SAME}} &= 0.236 \quad [0.203,\, 0.268], \\
\tau_{\text{resid,SAME}} &= 0.235 \quad [0.204,\, 0.267],
\end{align}
where brackets denote $95\%$ bootstrap CIs.  
By contrast, subliminal leakage to different-base students drops substantially:
\begin{align}
\tau_{\text{DIFF1}} &= 0.120 \quad [0.090,\, 0.154], \\
\tau_{\text{DIFF2}} &= 0.133 \quad [0.093,\, 0.170].
\end{align}
The near-equality of $\tau$ and $\tau_{\text{resid}}$ across all conditions confirms that public logits do not explain $y_{\text{priv}}$, validating that residualization correctly isolates the hidden channel (\S\ref{subsec:probes}).

\paragraph*{Alignment structure}
Despite consistently \emph{high global CKA} ($\geq 0.84$), only the same-base student exhibits substantial trait-subspace alignment:
\begin{align}
\mathrm{CKA}_{\text{sub}}(\text{SAME}) &= 0.630, \\
\mathrm{CKA}_{\text{sub}}(\text{DIFF1}) &= 0.241, \\
\mathrm{CKA}_{\text{sub}}(\text{DIFF2}) &= 0.207.
\end{align}
Thus, \textbf{global representational similarity is not diagnostic of covert-channel risk}; what matters is alignment along the narrow trait subspace, which same-base models inherit by construction. This empirically falsifies the assumption that global CKA is a sufficient proxy for subliminal-transfer capacity \cite{anthropic_subliminal2025}.

\begin{table}[t]
\centering
\caption{Seed effects: leakage and alignment.}
\label{tab:seed}
\setlength{\tabcolsep}{5pt}
\renewcommand{\arraystretch}{1.1}
\begin{tabular}{lcccc}
\toprule
Condition & Global CKA & Subspace CKA & $\tau$ & $\tau_{\text{resid}}$ \\
\midrule
SAME   & 0.979 & \textbf{0.630} & \textbf{0.236} & \textbf{0.235} \\
DIFF1  & 0.845 & 0.241 & 0.120 & 0.120 \\
DIFF2  & 0.922 & 0.207 & 0.133 & 0.131 \\
\bottomrule
\end{tabular}
\footnotesize\emph \\ Same-base student inherits trait-subspace alignment ($0.63$) and exhibits significant leakage ($\tau \approx 0.24$).  
DIFF-seed students retain high global similarity (0.84–0.92) but misalign in the trait subspace ($\approx 0.21$–0.24), collapsing to chance leakage ($\tau \approx 0.12$–0.13).
\end{table}

\subsection{Checkpoint \texorpdfstring{$\times$}{×} Data Ablation}
\label{subsec:ablation}

\begin{figure}[t]
\centering
\includegraphics[width=0.95\linewidth]{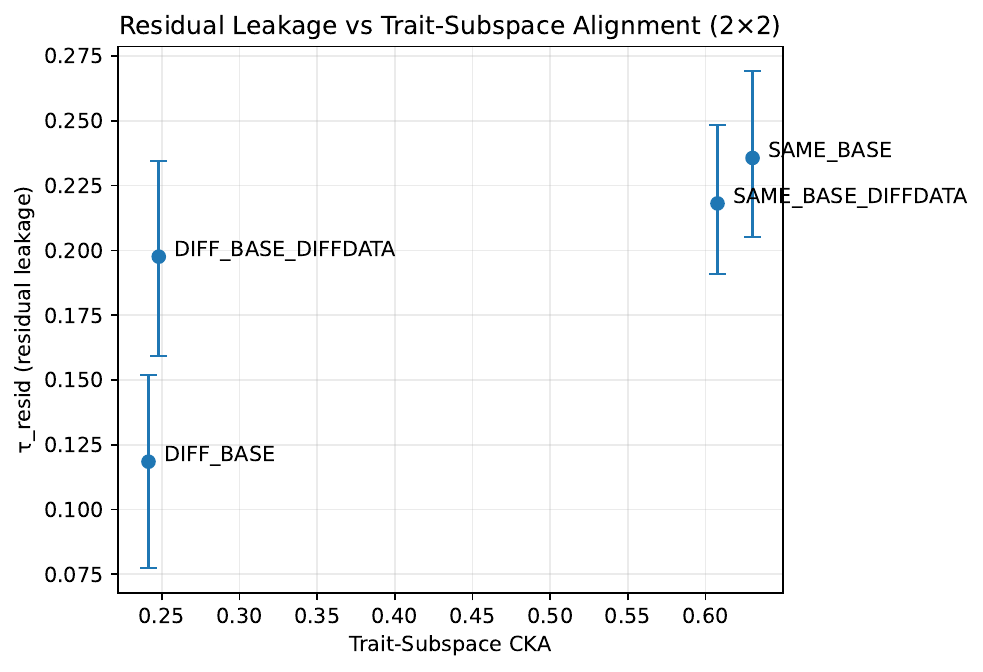}
\caption{$\tau_{\text{resid}}$ vs.\ subspace CKA across ablation conditions.}
\label{fig:tau_vs_subcka_clean}
\end{figure}

We next cross initialization (SAME\_BASE vs.\ DIFF\_BASE) with dataset (SAME vs.\ DIFFDATA). Table~\ref{tab:ablation} and Fig.~\ref{fig:tau_vs_subcka_clean} summarize the results.

\paragraph*{Initialization dominates}
SAME\_BASE and SAME\_BASE\_DIFFDATA both sustain strong trait-subspace CKA ($\approx 0.61$–$0.63$) and high leakage ($\tau \approx 0.22$–0.24), indicating that shared initialization is the critical enabler of subliminal transfer.  

\paragraph*{Global CKA is misleading}
Global CKA was computed on the teacher‑labeled validation features used for KD alignment; we observed that KD can inflate global alignment even under independent seeds. 
Global CKA can be inflated by distillation alignment on the public task, and therefore may not reflect trait‑subspace compatibility that governs subliminal transfer.
 This inflation does not extend to the trait‑subspace ($\mathrm{CKA}_{\text{sub}}\approx0.25$) and does not guarantee high leakage ($\tau=0.2$), reinforcing our subspace‑level analysis. This decoupling illustrates that global alignment induced by distillation on the public task does not imply trait‑subspace compatibility, underscoring that \emph{global} similarity alone is not predictive of covert-channel viability.

\begin{table}[t]
\centering
\caption{Checkpoint $\times$ Data ablation.}
\label{tab:ablation}
\setlength{\tabcolsep}{5pt}
\renewcommand{\arraystretch}{1.1}
\begin{tabular}{lcccc}
\toprule
Condition & Global CKA & Subspace CKA & $\tau$ & $\tau_{\text{resid}}$ \\
\midrule
SAME\_BASE           & 0.979 & \textbf{0.630} & \textbf{0.236} & \textbf{0.236} \\
SMBASE\_DFDATA & 0.975 & 0.608 & 0.218 & 0.218 \\
DFBASE           & 0.845 & 0.241 & 0.121 & 0.118 \\
DFBASE\_DFDATA & 0.974 & 0.248 & 0.200 & 0.198 \\
\bottomrule
\end{tabular}
\footnotesize\emph{Note:} Same-base models preserve trait-subspace alignment and leakage ($\tau\!\approx\!0.22$–$0.24$).  
Subspace CKA drops substantially ($\approx0.21$–0.24) and for different-base models despite global CKA exceeding $0.97$.
\end{table}

\subsection{Targeted Mitigation via Projection Penalty}
\label{subsec:mitigation-results}

We then evaluate a mitigation strategy that penalizes projection of student [CLS] activations onto the teacher’s trait subspace during KD (\S\ref{subsec:mitigation}):
\begin{equation}
\mathcal{L}_{\text{proj}}
= \alpha \cdot \mathbb{E}\!\left[\| U^{\top} \text{CLS}_{s} \|^{2}\right],
\qquad \alpha=10^{-2}.
\end{equation}

\paragraph*{Leakage suppression without task decay}
Relative to SAME\_BASE, projection penalty reduces alignment and leakage:
\begin{align}
\mathrm{CKA}_{\text{sub}}: & \; 0.630 \;\to\; 0.541, \\
\tau: & \; 0.236 \;\to\; 0.212, \\
\tau_{\text{resid}}: & \; 0.236 \;\to\; 0.214.
\end{align}
Meanwhile, public-task fidelity remains intact (public match $=1.000$ at every epoch), confirming that mitigation selectively disrupts covert-channel capacity without harming main-task performance.  

\subsection{Fidelity and Sanity Checks}
\label{subsec:fidelity}

\paragraph*{Public-task fidelity}  
Student public logits match teacher argmax on validation with $100\%$ accuracy across all epochs, ensuring that KD reproduced the intended task faithfully.  

\paragraph*{Teacher subliminal encoding}  
A probe on the teacher [CLS] yields $0.719$ private accuracy, confirming that the teacher reliably embeds the hidden trait \cite{anthropic_subliminal2025}.  

\paragraph*{Residualization validity}  
The close tracking of $\tau$ and $\tau_{\text{resid}}$ across conditions demonstrates that residualization removes public-logit predictability while isolating the hidden channel.

\subsection{Safety Implications}
\label{subsec:synthesis}

Results yield three key conclusions:

\begin{enumerate}
\item \textbf{Seed alignment governs subliminal transfer:} Independently initialized students resist covert-channel transfer despite global CKA $>0.9$, confirming that trait-subspace alignment—not global similarity—enables subliminal leakage.  

\item \textbf{Global CKA is an unreliable risk proxy:} High global overlap (e.g., DIFF\_BASE\_DIFFDATA, CKA$=0.974$) coexists with chance-level leakage, undermining its use as a security diagnostic.  

\item \textbf{Subspace-aware mitigation is effective:} Projection penalty lowers leakage by $\sim$0.02–0.03 absolute while preserving task fidelity, demonstrating a viable approach for hardening Transformer pipelines against covert inter-model communication.
\end{enumerate}

\section{Conclusion\label{sec:conc}}

We investigated the conditions under which subliminal transfer emerges between Transformer models and provided the first systematic evidence that \textbf{initialization seed alignment is the decisive enabler of covert channels}. 
While prior work assumed that high global representational similarity sufficed for subliminal transfer \cite{anthropic_subliminal2025}, our experiments demonstrate that independently initialized students (DIFF) fail to sustain hidden-signal transfer, even when global CKA exceeds $0.9$. 
Instead, successful transfer requires strong \emph{trait-subspace alignment}, which same-base models inherit by construction.

This finding reframes subliminal learning from an unavoidable property of shared architecture to a controllable property governed by initialization. 
In practice, this means that architectures deployed in federated or multi-agent systems may resist subliminal communication if initialized independently, despite converging to similar global solutions. 
We further showed that subspace-level CKA provides a diagnostic signal for covert-channel viability, overturning reliance on global metrics. 

Finally, we proposed and evaluated security controls, including a projection penalty that reduces trait-subspace alignment and suppresses leakage without impairing main-task accuracy. 
This result highlights the feasibility of \emph{subspace-aware defenses}, situating them alongside adversarial training and explanation regularization \cite{ross2017right} as tools to harden AI systems.

Deployment guidance for future AI systems to thwart subliminal transfer attacks: (i) prefer independently seeded replicas; (ii) monitor trait‑subspace CKA during model onboarding; (iii) apply projection penalties when white‑box teacher access exists. For scalable deployment, our projection penalty is preferable due to its simplicity and compatibility with encrypted inference. However, combined strategies--e.g., projection + adversarial reversal--may be warranted in high-assurance contexts that require stronger suppression. Particularly, modular AI stacks in avionics (e.g., perception, intent prediction, guidance) may run architecturally similar models across suppliers. Beyond avionics, independent seeding and subspace diagnostics also enhance privacy in federated healthcare models and prevent covert signaling in encrypted financial agents. Even though adversarial gradient reversal offers stronger suppression, it also increases compute costs and introduces optimization instability\cite{10839383}, limiting scalability and clashing with the assurance requirement. Right-for-the-wrong-reasons regularization similarly is sensitive to hyperparameter tuning even though it  enforces gradient orthogonality\cite{ross2017right}. Independent seeding across modules reduces the risk of subliminal inter‑module signaling, and subspace‑CKA provides an acceptance test during software integration prior to flight certification.

\paragraph*{Future Directions}  
Our study indicates that independently initialized Transformers can reduce subliminal transfer even when global representations appear similar, and that a subspace‑aware CKA diagnostic better tracks risk than global measures alone. While our evaluation uses a controlled synthetic corpus, the protocol is general and highlights practical levers—independent seeds and subspace‑aware regularization—for secure deployments. Future work should extend these diagnostics to larger models and real‑world tasks, but even now, our results provide actionable guidance for AI safety and cybersecurity contexts.

First, exploring whether the subspace misalignment property generalizes across larger models and natural datasets would test the limits of seed-induced uniqueness. 
Second, integrating subspace diagnostics into Continuous Integration and Continuous Deployment (CI/CD) pipelines could provide automated monitoring of covert-channel risk and detect emergent covert channels during model validation. Such deployment complements behavioral testing and aligns with NIST’s emphasis on measurable risk indicators\cite{autio_artificial_2024,sheh_cybersecurity_2024} and the EU AI Act’s requirement for technical robustness and traceability \cite{kilian_european_2025}. To further improve legal framework compatibility, nonlinear probes such as kernel methods \cite{white-etal-2021-non} that detect complex trait encodings as well as information theoretic metrics such as mutual information \cite{pimentel-etal-2020-information} and entropy-based leakage \cite{al-shehari_insider_2021} allowing broader validation across architectures and tasks can be used instead of our linear probes used in our example. These would support the EU AI Act's call for explainability and interpretability in high-risk systems\cite{sovrano_metrics_2022}.
Third, extending security controls to federated and encrypted Transformer settings \cite{zheng2023primer, moon2024thor, zhang2024nexus} may yield stronger resilience in adversarial environments. 
Together, these steps will help refine the security model for resilient, auditable, and regulation-ready collaborative AI systems, ensuring that shared architectures do not silently enable subliminal communication.

\section*{ACKNOWLEDGMENT}
This material is based upon work supported by the National Science Foundation award CNS-2244515 and the Embry-Riddle Aeronautical University Office of Undergraduate Research. 
Portions of this manuscript were augmented using Microsoft 365 Copilot Researcher and Writing Coach Agents. The final content was reviewed and confirmed by the authors.

\bibliographystyle{jabbrv_IEEEtran}
\bibliography{references}
    
\end{document}